\pdfoutput=1
\documentclass[11pt]{arxiv-article}
	
\bibliography{References}

\usepackage{Definitions}
\usepackage{tikz}

\usepackage{pifont}
\usepackage{hyperref}
\usepackage{mathrsfs}  

\def\mathclap#1{\text{\hbox to 0pt{\hss$\mathsurround=0pt#1$\hss}}}

\allowdisplaybreaks

\newcommand{\OfficialTitle}{
  Near-Schrödinger dynamics at large charge
}
\title{\setstretch{1.4}
  {\color{Thoughtless}\Huge\textbf{\dosserif\OfficialTitle}}
}

\hypersetup{pdfauthor={},pdftitle={\OfficialTitle}}

\author{%
  \begin{minipage}{.97\linewidth}
    \vspace{1cm}
    \begin{center} \dosserif%
      {\small
        \textbf{Domenico~Orlando}\textsuperscript{\ding{72}\ding{73}},
        \textbf{Vito~Pellizzani}\textsuperscript{\ding{73}},
         and
         \textbf{Susanne~Reffert}\textsuperscript{\ding{73}} 
         }
    \end{center}
    \vspace{1cm}
    \authorBlock{\ding{72}}{\dosserif{} INFN sezione di Torino | Arnold--Regge Center\\
      via Pietro Giuria 1, 10125 Turin, Italy}
    \authorBlock{\ding{73}}{\dosserif{} Albert Einstein Center for Fundamental Physics\\
      Institute for Theoretical Physics, University of Bern,\\
      Sidlerstrasse 5, CH-3012 Bern, Switzerland}
  \end{minipage}
}

\date{}

\begin{document}

\setstretch{1.2}

\numberwithin{equation}{section}

\begin{titlepage}

  \newgeometry{top=23.1mm,bottom=46.1mm,left=34.6mm,right=34.6mm}

  \maketitle

  \thispagestyle{empty}

  \vfill\dosserif{}

  \abstract{\normalfont{}\noindent{}%
    In this note we discuss a non-relativistic system at large charge in a regime where Schrödinger symmetry is slightly broken by an explicit mass term for the dilaton field which non-linearly realizes non-relativistic scale invariance.
    To get there, we first develop the large-charge formalism from the linear sigma model perspective, including the harmonic trapping potential necessary for the non-relativistic  state-operator correspondence.
    As a signature of the explicit breaking, we identify a $\sqrt{Q}\log{Q}$ term, which depending on the space dimension is either of the same order as the effects coming from the breakdown of the \acs{eft} at the edge of the particle cloud, or can be distinguished from these effects over a large range of orders of magnitude.
  }

\vfill

\end{titlepage}

\restoregeometry{}

\setstretch{1.2}

\tableofcontents

\section{Introduction}%
\label{sec:introduction}

Working in a sector of large global charge has in recent years proven to be a powerful tool to access strongly coupled systems which were previously inaccessible to analytic methods~\cite{Hellerman:2015nra,Alvarez-Gaume:2016vff,Monin:2016jmo,Banerjee:2019jpw,Gaume:2020bmp}. The large charge appears as a controlling parameter in a perturbative expansion, allowing the calculation of physical observables. The spontaneous symmetry breaking due to the classical ground state at large charge gives rise to Goldstone \acp{dof} in terms of which the effective action governing the low-energy physics can be expressed.

This approach has been used mostly in the context of \acp{cft}, where the space-time symmetry gives rise to strong constraints on the form of the correlators themselves and the terms appearing in the effective action. The strength of working at large charge lies in the fact that the Wilsonian effective action compatible with the symmetries can be truncated after a few terms, since further terms are suppressed by inverse powers of the large charge. The state-operator correspondence of \ac{cft}~\cite{Cardy:1984rp,Cardy:1985lth} moreover simplifies the calculation of the operator dimension of the lowest operator of a given charge $Q$, which corresponds to the energy of the ground state in the sector of fixed charge $Q$. 

A first attempt to go beyond the conformal regime was made in~\cite{Orlando:2019skh,Orlando:2020yii}, where Coleman's dilaton dressing~\cite{Coleman:1988aos} was used as a mechanism to explicitly break conformality in a controlled way by giving the dilaton field a {(fine-tuned)} small mass term.

\medskip
It has been shown that also non-relativistic systems with Schrödinger symmetry (also called non-relativistic \acp{cft}), realized in the lab by the unitary Fermi gas~\cite{randeria2012bcs},  lend themselves to the treatment at large charge~\cite{schroedinger,Kravec:2018qnu,Kravec:2019djc}.\footnote{For related work, see also~\cite{Arav:2017plg,Argurio:2020jcq}.} While the constraints on the correlators are less stringent than in \ac{cft}, the effective action at large charge can be written down following the same approach as for \acp{cft}, combining the constraints from Schrödinger symmetry and the large-charge scaling.
Also, the state-operator correspondence has its analog in non-relativistic systems, where the role of the cylinder frame is played by a harmonic potential of strength \(\omega\), which traps the particle cloud~\cite{Nishida:2007pj, Goldberger:2014hca}, so that the conformal dimensions in the plane are proportional to the energies in the trap:
\begin{equation}
  \Delta = \frac{E}{\hbar \omega} .
\end{equation}

\medskip
In this note, we aim to extend the large-charge approach to systems with near-Schrödinger dynamics, following the path marked out by Coleman. We proceed analogously to~\cite{Orlando:2019skh}, by first non-linearly realizing Schrödinger symmetry in a system with Galilean symmetry via a dilaton dressing. 

{When we consider our system on a compact space (e.g. on the torus) or in a trapping potential, fixing the charge gives rise to spontaneous symmetry breaking and the low-energy physics is encoded by the corresponding Goldstone boson, even if the original system is strongly coupled.} {If we work on the torus (but without a trapping potential),} we can start from a Lagrangian with Galilean symmetry  of the form
\begin{equation}
	\mathcal{L} = -k_0 + k_1 \left( \dot{\chi} - \frac{\hbar}{2m} (\del_i \chi)^2 \right) + \dots,
\end{equation}
where $\chi$ is a Goldstone boson and dress operators of dimension $k$ using the newly introduced dilaton field $\sigma$:
\begin{equation}
	\mathcal{O}_k(t, x_i) \to e^{\frac{2 (k - d - 2)}{d} f \sigma(t, x_i)} \mathcal{O}_k(t, x_i).
\end{equation}
This leads to the Schrödinger-invariant action 
\begin{equation} 
	\mathcal{L}(\psi) = c_0 \left[ \frac{i}{2} ( \psi^*\del_t \psi - \psi \del_t \psi^*) - \frac{\hbar}{2m} \del_i \psi^* \del_i \psi \right] - \frac{\hbar c_1}{2 m} e^{-2 f \sigma} (\del_i \sigma)^2 - \frac{\hbar c_2}{m \hbar^\frac{2}{d}} (\psi^*\psi)^\frac{d+2}{d}
\end{equation}
in terms of the complex field $\psi = \frac{1}{f} e^{- f \sigma - i \chi}.$

Since we want to keep working in the \ac{lsm}, we rederive the large-charge results of the Schrödinger system in the harmonic trap of~\cite{Kravec:2018qnu} in this formalism, where we also keep track of \ac{nlo} terms. 
We find for the operator dimension of the lowest state of charge $Q$
\begin{equation}
\Delta =
\begin{cases}
	\frac{2}{3} \xi Q^\frac{3}{2} + \mathcal{O}(Q^\frac{1}{2} \log Q)  & (d = 2) \\
	\frac{3}{4} \xi Q^\frac{4}{3} + \frac{27}{8 \xi} \frac{c_1}{c_0}  Q^\frac{2}{3} + \mathcal{O}(Q^\frac{5}{9}) & (d = 3),
\end{cases}
\end{equation}
where $\xi$ is a constant of order 1.
We additionally compute the Casimir energy of the fluctuations over the ground state which in $d=2$ is given by
\begin{equation}
  E^{(d=2)}_{\text{Cas}} = -0.294159\dots \times \omega.
\end{equation}

{Next, we want to investigate a small departure from Schrödinger symmetry. The mechanism we use consists in giving the \ac{eft} dilaton a small mass \(m_\sigma\).} We calculate the ground-state energy in presence of this breaking term and the correction to the scaling dimension of the lowest operator with charge $Q$:
\begin{equation}
\tilde{\Delta} = 
\begin{cases}
	\frac{2}{3} \xi Q^\frac{3}{2} + \mathcal{O}(Q^\frac{1}{2} \log Q)  & (d = 2) \\
	\frac{3}{4} \tilde{\xi} Q^\frac{4}{3} + \frac{27}{8 \tilde{\xi}} \frac{c_1}{c_0}  Q^\frac{2}{3} - \frac{\kappa m_\sigma^2}{\tilde{\xi}^{3/2}} \sqrt{Q} \log(Q) + \mathcal{O}(Q^{\frac{5}{9}}) & (d = 3),
\end{cases}
\end{equation}
where \(\kappa\) is an order-one coefficient.

One of the big stumbling blocks we encounter is our ignorance of the contributions of the edge of the particle cloud in the harmonic potential to the effective action, forcing us to work with estimates instead. In fact, in $d=2$, the signature of the dilaton mass is of the same order as the uncertainty due to the cloud edge.
The edge operators appearing in the \ac{eft} have been addressed in~\cite{Hellerman:2020eff}, but alas too late to be of use to us in this work.
A discussion of our problem based on the edge \ac{eft} should be attempted in the future.%

\bigskip

This note is organized as follows. In Section~\ref{sec:schroedinger}, we discuss the dilaton dressing to non-linearly realize the full Schrödinger symmetry in a system with Galilean invariance. In Section~\ref{sec:LQschroedinger}, we rederive the results on Schrödinger-invariant systems at large charge from~\cite{schroedinger,Kravec:2018qnu} in the \ac{lsm} to set the stage for the near-Schrödinger case. Based on the interpretation of the radial mode as a dilaton, we then add a small explicit mass term for the dilaton which explicitly breaks Schrödinger symmetry and calculate the corrections arising from this breaking to the large-charge results (Section~\ref{sec:near-schroedinger}). In Section~\ref{sec:conclusions}, we discuss our results and further research directions.

\section{Non-linear realization of Schrödinger symmetry}%
\label{sec:schroedinger}

We consider a Schrödinger-symmetric system with a complex scalar field in $d+1$ dimensions and a global U(1) symmetry that we interpret as the particle number.
We know that in a sector of fixed charge, the ground state spontaneously breaks the global symmetry, giving rise to a Goldstone field $\chi$ in terms of which we want to write down an effective action. 

We proceed in analogy to Coleman~\cite{Coleman:1988aos}, starting with a quadratic, Galilean-invariant action for the Goldstone $\chi$. We then non-linearly realize the full Schrödinger symmetry by introducing a dilaton field and dressing appropriately all the operators in the Lagrangian.
To our knowledge, the closest attempt in this direction was made in~\cite{Arav:2017plg}. More recently, also~\cite{Argurio:2020jcq} suggested the same idea.
 
\medskip
The Galilean algebra is generated by (time and space) translations, rotations and Galilean boosts (uniform motion with velocity $\vec v$). It can be centrally extended by the particle number generator. Together with the non-relativistic scale translation,
\begin{equation}\label{eq:nonrela_dila_transfo}
(t,x_i)  \rightarrow (t',x_i') = (e^{2\tau}t,e^{\tau} x_i),
\end{equation}
where $\tau$ is a real parameter, and non-relativistic \ac{sct}, 
\begin{equation}\label{eq:nonrela_SCT} 
(t,x_i)  \rightarrow (t', x_i') = \left(\frac{t}{1 + \lambda t}, \frac{x_i}{1 + \lambda t}\right),
\end{equation}
 with $\lambda$ a real parameter, it forms the Schrödinger algebra (for more details see e.g. the appendix of~\cite{schroedinger}).

\medskip   
{As explained in the introduction, the physics at fixed charge is described by a Goldstone boson $\chi$.} We start out with the most general non-relativistic action for $\chi$ invariant under Galilean symmetry
expressed as function of the quantity
\begin{equation}
	U = \del_t \chi - \frac{\hbar}{2m} \del_i \chi \del_i \chi,
\end{equation}
where $m$ is a mass parameter. This function $F(U)$ can be expanded in a Taylor series. The first terms read
\begin{equation} \label{eq:GalileanLag}
\begin{aligned}
	F(U) &= -k_0 + k_1 U + k_2 U^2 + …\\
	 &= -k_0 + k_1 \left( \dot{\chi} - \frac{\hbar}{2m} (\del_i \chi)^2 \right) + k_2 \left( \dot{\chi}^2-\frac{\hbar}{m} \dot{\chi} (\del_i \chi)^2 + \frac{\hbar^2}{4m^2} (\del_i \chi)^4 \right) + \dots .
\end{aligned}
\end{equation}
{As discussed in~\cite{schroedinger,Kravec:2018qnu} (see also~\cite{Son:2005rv}), in the large-charge limit we can limit ourselves to the $k_0$- and $k_1$-terms, since the higher-order terms are parametrically smaller.}
Note that the coefficients are dimensionful: 
\begin{equation}
	[k_i] = M L^{2-d} T^{i-2}.
\end{equation}
In order to promote this Lagrangian to a scale-invariant one, we introduce (in parallel to the construction of Coleman in the relativistic case~\cite{Coleman:1988aos}) a new field $\sigma$ -- the dilaton -- which transforms non-linearly under non-relativistic scale transformations~\eqref{eq:nonrela_dila_transfo} as
\begin{equation}
	\sigma(t, x_i)  \rightarrow \sigma(t, x_i) + \frac{d}{2 f} \tau,
\end{equation}
where $f$ is a constant with units $[f^{-1}] = [\sigma] = M^\frac{1}{2} L^\frac{2-d}{2} T^{-\frac{1}{2}}$. Any operator $\mathcal{O}_k$ of dimension $k$ can be dressed with an appropriate power of $\sigma$ to become scale invariant:
\begin{equation}
	\mathcal{O}_k(t, x_i) \to e^{\frac{2 (k - d - 2)}{d} f \sigma(t, x_i)} \mathcal{O}_k(t, x_i).
\end{equation}
In particular, the constant operator should be dressed as $k_0 \to k_0 e^{-\frac{2(d+2)}{d} f \sigma}$. The $k_1$-term has engineering dimension $2$. Hence, a fully Schrödinger-invariant Lagrangian has the structure
\begin{equation}
\begin{aligned}
	\mathcal{L} & = -\frac{\hbar \kappa}{2 m} e^{-2 f \sigma} (\del_i \sigma)^2 + e^{-\frac{2(d+2)}{d} f \sigma} F(e^{\frac{4}{d} f \sigma} U) \\
	& = -\frac{\hbar \kappa}{2 m} e^{-2 f \sigma} (\del_i \sigma)^2 - k_0 e^{-\frac{2(d+2)}{d} f \sigma} + k_1 e^{-2 f \sigma} U + ...
\end{aligned}
\end{equation}
Note that we have included a kinetic term for the dilaton (together with a new coefficient $\kappa$ with $[\kappa] = [k_1]$) that respects Schrödinger invariance.
We do not include time derivatives of $\sigma$ as they would break boost invariance~\cite{Arav:2017plg}.
It is convenient to combine the fields $\sigma$ and $\chi$ into a complex field\footnote{Note that this definition excludes $\psi = 0$, which is not an issue when the field $\psi$ is considered a small fluctuation around a non-zero \ac{vev}. This scenario, in turn, corresponds to turning on a non-zero charge density, which is what we are interested in.}
\begin{equation}
	\psi = \frac{1}{f} e^{- f \sigma - i \chi}.
\end{equation}
In terms of $\psi$ the Lagrangian becomes
\begin{equation} \label{eq:Schroedinger_lag}
	\mathcal{L}(\psi) = c_0 \left[ \frac{i}{2} ( \psi^*\del_t \psi - \psi \del_t \psi^*) - \frac{\hbar}{2m} \del_i \psi^* \del_i \psi \right] - \frac{\hbar c_1}{2 m} e^{-2 f \sigma} (\del_i \sigma)^2 - \frac{\hbar c_2}{m \hbar^\frac{2}{d}} (\psi^*\psi)^{\frac{2}{d}+1},
\end{equation}
where the dimensionless (Wilsonian) coefficients $c_i$ depend on the previous coefficients through 
\begin{align}
	c_0 &= f^2 k_1, & c_1 &= \kappa - f^2 k_1, & c_2 &= \frac{m}{\hbar^2} (\hbar f^2)^{\frac{2}{d}+1} k_0.
\end{align}
Higher-derivative terms would correspond to allowing higher-order terms in Eq.~\eqref{eq:GalileanLag}. { This Lagrangian is fully symmetric under the Schrödinger group, including special conformal transformations Eq.~\eqref{eq:nonrela_SCT}, provided that the fields transform as~\cite{Arav:2017plg, schroedinger}
\begin{equation}
\begin{cases}
	\sigma(t, x_i) & \stackrel{SCT}{\longrightarrow}~ \sigma(t, x_i) - \frac{d}{2 f} \ln(1 + \lambda t) \\
	\chi(t, x_i) & \stackrel{SCT}{\longrightarrow}~ \chi(t, x_i) + \frac{m}{2 \hbar} \frac{\lambda \vec x^2}{1 + \lambda t}.
\end{cases}	
\end{equation}
This yields the usual transformation law for $\psi$ in the Schrödinger model \cite{Hagen:1972pd, schroedinger}. Hence, $\mathcal{L}$ coincides with the Schrödinger Lagrangian together with a kinetic term for the dilaton. Note that invariance of the $c_0$-term is ensured by the fact that the time derivative transforms non-trivially under Eq.~\eqref{eq:nonrela_SCT}: $\del_t \to (1 + \lambda t)^2 \del_t + \lambda (1 + \lambda t) x_i \del_i$. Indeed, it decomposes into
\begin{equation}
\begin{aligned}
		\frac{i}{2} (\psi^* \dot\psi - \psi \dot\psi^*) \to (1 + \lambda t)^{d + 2} & \left[ \frac{i}{2} (\psi^* \dot\psi - \psi \dot\psi^*) + \frac{m}{2 \hbar} \frac{\lambda^2 \vec x^2}{(1 + \lambda t)^2} \psi^* \psi \right. \\
		& \quad \left. + \frac{i \lambda x_i}{2 (1 + \lambda t)} (\psi^* \del_i \psi - \psi \del_i \psi^*) \right]
\end{aligned}
\end{equation}
and
\begin{equation}
\begin{aligned}
		\frac{\hbar}{2m} |\del_i \psi|^2 \to (1 + \lambda t)^{d+2} & \left[ \frac{\hbar}{2m} |\del_i \psi|^2 + \frac{m}{2 \hbar} \frac{\lambda^2 \vec x^2}{(1 + \lambda t)^2} \psi^* \psi \right. \\
		& \quad \left. + \frac{i \lambda x_i}{2 (1 + \lambda t)} (\psi^* \del_i \psi - \psi \del_i \psi^*) \right].
\end{aligned}
\end{equation}
Subtracting the second line from the first shows the invariance of this term under \ac{sct}. The last two terms of Eq.~\eqref{eq:Schroedinger_lag} are readily seen to be invariant as they do not contain time derivatives. For instance, invariance of the $c_1$-term is shown as follows:
\begin{equation}
		e^{-2 f \sigma} (\del_i \sigma)^2 \to e^{-2 f \sigma + d \ln(1 + \lambda t)} [(1 + \lambda t) \del_i \sigma]^2 = (1 + \lambda t)^{d+2}  e^{-2 f \sigma} (\del_i \sigma)^2.
\end{equation}}

Without the kinetic term of the dilaton, integrating out the radial mode would lead to a leading-order \ac{nlsm} Lagrangian of the form $U^{\frac{d+2}{2}}$. We expect the kinetic term of the dilaton to have the effect of giving an \ac{nlo} correction. Once we turn on a harmonic trapping potential, we will have to distinguish the semiclassical (bulk) contributions from the effects coming from the boundary of the particle cloud and have to verify that this term is indeed subleading.

\section{The Schrödinger particle at large charge}%
\label{sec:LQschroedinger}

Despite the fact that time and space scale differently in the non-relativistic scale transformation Eq.~\eqref{eq:nonrela_dila_transfo}, a number of analogies between the relativistic and non-relativistic cases persist. The relativistic O(2) vector model with its global U(1) symmetry is the simplest model one can study at large charge~\cite{Hellerman:2015nra,Gaume:2020bmp}, and our treatment of the non-relativistic Schrödinger system follows along the same lines~\cite{schroedinger,Kravec:2018qnu}. Let us first recall some well-known results.

We are interested in the conformal dimension of the lowest operator with fixed charge $Q \gg 1$ corresponding to the U(1) generator of the particle number and will compute it using the state-operator correspondence.

In the relativistic case, this correspondence is based on the fact that $\mathbb{R}^{d+1}$ (flat space) and $\mathbb{R} \times S^d(R_0)$ (cylinder) are related by a Weyl transformation and therefore conformally equivalent.
The conformal dimensions of operators inserted in flat space are identified with the energies of states on the sphere $S^{d}$ via
\begin{equation}\label{eq:state-opDeltaE}
  \Delta = R_0 E_{S^d}.
\end{equation}
The radius of the sphere $R_0$ sets an \ac{ir} cutoff.
In the $O(2)$ model at fixed charge, the global symmetry is spontaneously broken and the radial mode becomes massive and decouples for energies smaller than $Q^\frac{1}{d}/R_{0}$. The \ac{eft} is thus valid for energy scales $\Lambda$ in the range
\begin{equation}
	\frac{1}{R_0} \ll \Lambda \ll \frac{Q^\frac{1}{d}}{R_{0}},
\end{equation}
which in turn requires $Q \gg 1$. The natural expansion parameter is therefore $Q^{-\frac{1}{d}}$.

In Schrödinger systems, one can access the conformal dimension of an operator in a similar fashion.
In this case, the corresponding \ac{ir} cutoff is realized by trapping  the system in a spherical harmonic potential~\cite{Nishida:2007pj, Goldberger:2014hca},
\begin{equation}\label{eq:harmonicPot}
	A_0(\vec{x}) = \frac{m \omega^2}{2 \hbar} r^2,
\end{equation}
where $r = \abs{\vec{x}}$ and $\omega > 0$ defines the strength of the potential.
The energy spectrum of the trapped system is isomorphic to the set of conformal dimensions of operators inserted in flat space without the trap, \emph{i.e.}
\begin{equation}
	\Delta = \frac{1}{\hbar \omega} E_{harm}.
\end{equation}
The harmonic potential confines the particles in a spherical \emph{cloud} at the edge of which the charge density falls rapidly to zero. The bulk \ac{eft} description is thus limited by our ignorance of what happens in this region where quantum effects become important in the form of \ac{ir} divergences that need to be regularized.
This naturally sets an \ac{ir} length-scale cutoff called the \emph{cloud radius},
\begin{equation}\label{eq:cloudRadius}
	R_{\text{cl}} = \sqrt{\frac{2 \hbar \mu}{m \omega^2}},
\end{equation}
which measures the distance from the center of the cloud to where the semiclassical number density vanishes.
$\mu$ is a parameter that will appear in the ground-state solution at fixed charge and is interpreted as a charge-dependent chemical potential \( \mu \approx Q^{1/d}\).
The fixed-charge \ac{eft} also has an \ac{uv} cutoff associated with the breaking of the Schrödinger symmetry.
This sets the scale beyond which the gapped radial mode decouples, namely 
\begin{equation}\label{eq:Rmu}
	R_\mu = \sqrt{\frac{2 \hbar}{m \mu}}.
\end{equation}
The \ac{eft} is therefore valid in a regime where
\begin{equation} \label{eq:NR_scaling}
	R_{\text{cl}} \gg r \gg R_\mu = \frac{\omega}{\mu} R_{\text{cl}},
\end{equation}
which requires $\mu \gg \omega$. Hence, the natural expansion parameter is $\epsilon = \frac{\omega}{\mu}$ and this will turn out to be $\sim Q^{-\frac{1}{d}}$ (to leading order), just as before. Remarkably, this leads to the \emph{same} leading-order dependence on $Q$ for the conformal dimension of the lowest operator at large charge as in the relativistic case:
\begin{equation}
	\Delta \sim Q^\frac{d+1}{d}.
\end{equation}
One of the main differences to the relativistic case, though, is the \emph{explicit space-dependence} of the ground-state solution due to the harmonic trap.
Crucially, the bulk \ac{eft} description breaks down near the edge of the cloud, where the density of the particle cloud falls off. This effect should be compensated by a boundary \ac{eft} which captures contributions from the edge states~\cite{Hellerman:2020eff}. In absence of an effective boundary Lagrangian, we have to rely on estimates of the contribution of the droplet edge, following~\cite{Son:2005rv}. 

\subsection{Semiclassical results}%
\label{sec:semiclass}

Let us now turn to explicit computations. In view of adding a mass term for the dilaton later on, we work in the \ac{lsm} which explicitly contains the massive mode. We thus consider the Lagrangian~(\ref{eq:Schroedinger_lag}) coupled to the harmonic potential,
\begin{equation} \label{eq:lsm_harmonic}
	\mathcal{L}(\psi) = c_0 \left[ \frac{i}{2} \left( \psi^* (D_t \psi) - \psi (D_t \psi)^* \right) - \frac{\hbar}{2m} \abs{\del_i \psi}^2 \right] - \frac{\hbar c_1}{2 m} e^{-2 f \sigma} (\del_i \sigma)^2 - \frac{\hbar c_2}{m \hbar^\frac{2}{d}} (\psi^*\psi)^{\frac{2}{d}+1},
\end{equation}
where the harmonic potential appears through the covariant time-derivative $D_t = \del_t - i A_0(\vec{x})$. The charge density corresponding to the global U(1) of the particle number generator is given by
\begin{equation}
	\rho = \frac{c_0}{\hbar} \psi^* \psi
\end{equation}
and has units of $L^{-d}$. This makes the total charge dimensionless as it should be.

\medskip
As already mentioned, the radial mode $a(t,\mathbf{x}) = \abs{\psi(t,\mathbf{x})} = \frac{1}{f} e^{-f \sigma(t,\mathbf{x})}$ decouples when the Schrödinger symmetry is broken in presence of a fixed charge density, as the last term in Eq.~(\ref{eq:lsm_harmonic}) gives a mass term (that in our conventions has dimension $T^{-1/2}$) for the fluctuations.
This will be developed in detail in the next section, but we can already think in terms of the corresponding cutoff $R_\mu$. As it turns out, working in the regime Eq.~(\ref{eq:NR_scaling}), the dynamics of the massless mode $\chi$ corresponds to the description of a non-relativistic superfluid in a harmonic trap~\cite{Son:2005rv,Kravec:2018qnu}. It is most convenient to decompose the field into a radial and an angular mode, $\psi(t,\mathbf{x}) = a(t,\mathbf{x}) e^{-i \chi(t,\mathbf{x})}$. In this notation,
\begin{equation}\label{eq:LSMLagU}
	\mathcal{L}(\chi, a) = c_0 a^2 U - \frac{c_1 \hbar}{2 m} (\partial_i a)^2 - \frac{\hbar c_2}{m \hbar^\frac{2}{d}} a^{\frac{4}{d}+2},
\end{equation}
where the time derivative in $U$ was also promoted to a covariant derivative $D_t \chi = \del_t \chi - A_0(\vec{x})$ so that
\begin{equation} \label{eq:U_def}
	U(t,\mathbf{x}) = D_t \chi - \frac{\hbar}{2 m} (\partial_i \chi)^2.
\end{equation}
For later convenience, we replace from now on the dimensionless coupling $c_2$ by $g \equiv \left( \frac{4}{d} + 2 \right)  c_2$, so that the \ac{eom} read
\begin{equation}\label{eq:eom}
\begin{cases}
	a(t,\mathbf{x})^{\frac{4}{d}+1} & = \frac{c_0 \hbar^\frac{2}{d}}{g} \left[ \frac{2 m}{\hbar} a(t,\mathbf{x}) U(t,\mathbf{x}) + \frac{c_1}{c_0} \nabla^2 a(t,\mathbf{x}) \right], \\
	0 & = \del_t \rho(t,\mathbf{x}) + \del_i j_i(t,\mathbf{x}),
\end{cases}
\end{equation}
where
\begin{align}
	\rho(t,\mathbf{x}) &= \frac{c_0}{\hbar} a^2, & j_i(t,\mathbf{x}) &= -\frac{\hbar}{m} \rho  \del_i \chi.
\end{align}
The second line is the equivalent of the continuity equation for the superfluid. 
 Finally, the Hamiltonian density is given by
\begin{equation} \label{energy_density}
	\mathcal{E} = \frac{\hbar}{2 m} \left[ c_0 a^2 (\partial_i \chi)^2 + \frac{2 m c_0}{\hbar} A_0(\vec{x}) a^2 + c_1 (\partial_i a)^2 + \frac{d g}{(d + 2) \hbar^\frac{2}{d}} a^{\frac{4}{d}+2} \right].
\end{equation}

\subsubsection{Ground-state solution and scales}

Since the only \ac{dof} of our system are those of a complex scalar, we do not have enough of them to also account for Goldstone bosons arising from breaking further spatial symmetries beyond the global U(1).\footnote{This explains the richer zoology of ground states in model with non-Abelian symmetry, as discussed \emph{e.g.} in~\cite{Hellerman:2017efx,Hellerman:2018sjf,Banerjee:2019jpw}.} The ground-state solution must thus be spherically symmetric.
The simplest solution to the \ac{eom} has the form
\begin{equation}
  \begin{cases}
    a(t,\mathbf{x}) = v(r) \\
    \chi(t, \mathbf{x}) = \mu t,
  \end{cases}
\end{equation}
where \(v\) and \(\mu\) satisfy
\begin{equation} \label{eq:GS_EoM}
  \begin{cases}
    v(r)^{\frac{4}{d}+1} = \frac{c_0 \hbar^\frac{2}{d}}{g} \left[ \frac{2 m}{\hbar} v(r)  U_0(r) + \frac{c_1}{c_0}  \left( \del_r^2 v(r) + \frac{d - 1}{r} \del_r v(r) \right) \right], \\
    Q = \int \dd[d]{\mathbf{x}} \rho_0(r) = \frac{c_0}{\hbar} \int \dd[d]{\mathbf{x}} v(r)^2,
  \end{cases}
\end{equation}
and \(U_0(r)\) is the \ac{vev} of $U$ given in Eq.~\eqref{eq:U_def}, namely
\begin{equation}
  U_0(r) = \mu - \frac{m \omega^2}{2 \hbar} r^2 . 
\end{equation}
As explained above, the charge density \(\rho_0(r)\) has a non-trivial dependence on the distance from the origin due to the harmonic potential.
However, the part of the \ac{eom} that includes the Laplacian is a subleading contribution, which means that at leading order, the charge density $\rho_0(r)$  vanishes when $U_0(r) = 0$, as dictated by Eq.~(\ref{eq:GS_EoM}). The distance where this occurs is called the {\em cloud radius} \(R_{\text{cl}}\):
\begin{equation}
  U_0(R_{\text{cl}}) = 0 \quad \Rightarrow \quad \rho_0(R_{\text{cl}}) = 0 \quad \Rightarrow \quad R_{\text{cl}} = \sqrt{\frac{2 \hbar \mu}{m \omega^2} }.
\end{equation}
It is convenient to rescale the distances with respect to \(R_{\text{cl}}\) and define the adimensional quantity
\begin{equation}
  s = \frac{r^2}{R_{\text{cl}}^2}.
\end{equation}
The boundary of the cloud corresponds to \(s = 1\).
In these terms, the measure for spherically-invariant functions becomes
\begin{equation}
	\Omega_d r^{d-1} dr = \frac{\pi^\frac{d}{2} R_{\text{cl}}^d}{\Gamma\left(\frac{d}{2}\right)}  s^{\frac{d}{2}-1} ds,
\end{equation}
where $\Omega_ d = \frac{2 \pi^\frac{d}{2}}{\Gamma\left(\frac{d}{2}\right)}$ is the surface of a unit $d$-sphere. 

\bigskip

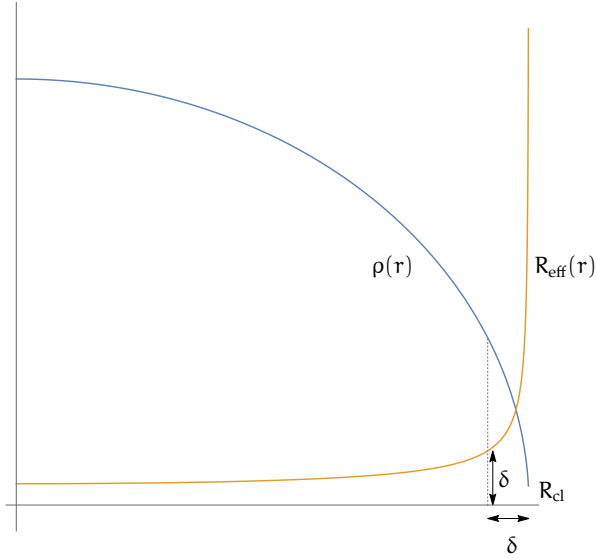
\begin{figure}
  \centering
  \begin{footnotesize}
  \begin{tikzpicture}[scale=.7]
    \node at (0,0) {\includegraphics[width=.5\textwidth]{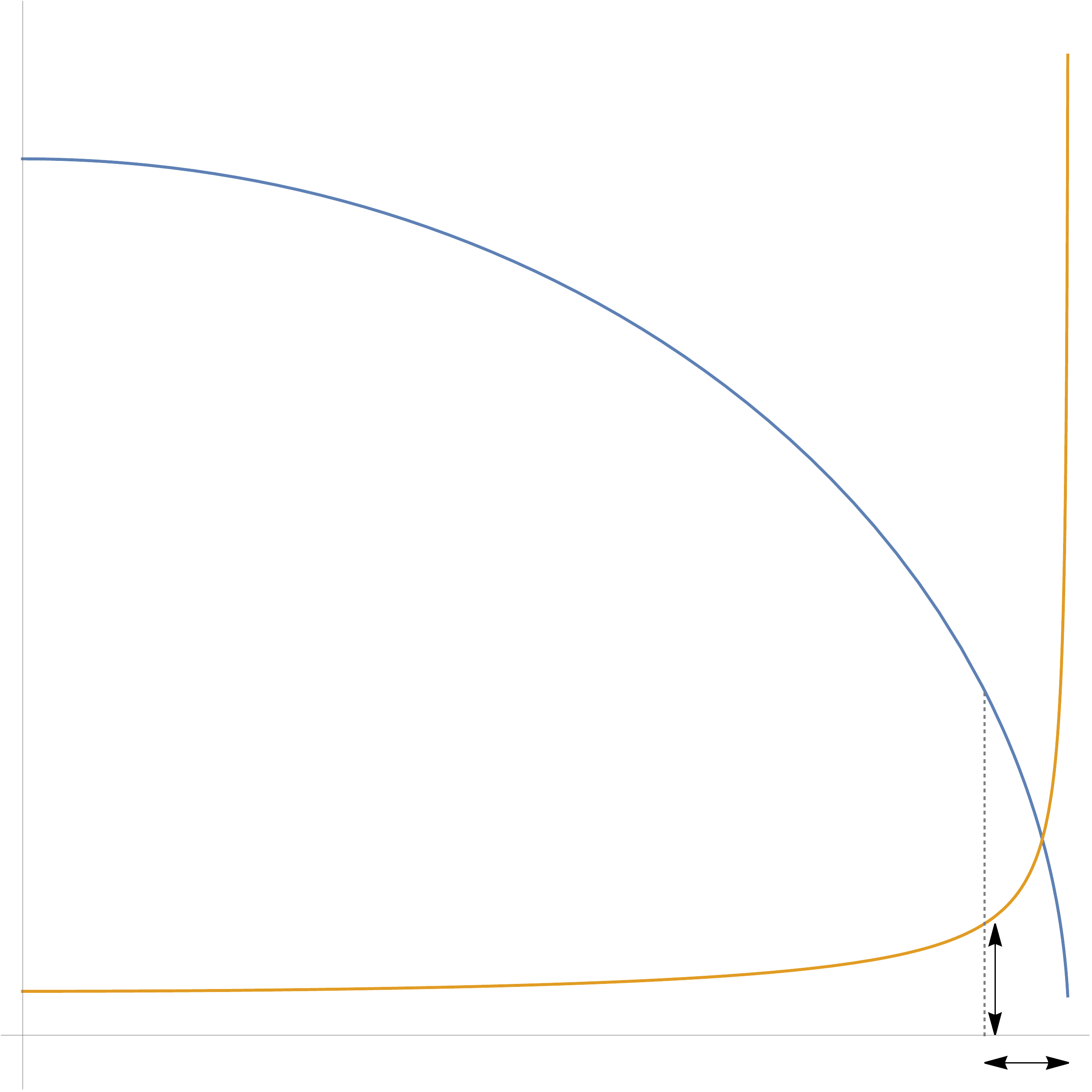}};
    \node at (2.25,0) {\(\rho(r)\)};
    \node at (5.5,0) {\(R_{\text{eff}}(r)\)};
    \node at (5.25,-4.25) {\(R_{\text{cl}}\)};
    \node at (4.35,-4) {\(\delta\)};
    \node at (4.5,-5.25) {\(\delta\)};
  \end{tikzpicture}
\end{footnotesize}
  \caption{Scales \(R_{\text{cl}}\) and \(\delta\). The blue line represents the charge density as a function of \(r\), the orange line represents the effective length \(R_{\text{eff}}\) that bounds the validity of the EFT description. The distance \(\delta\) marks the point where the scale that we want to probe becomes of the same order as \(R_{\text{eff}}\) and the EFT cannot be trusted anymore.}
  \label{fig:scales}
\end{figure}

The existence of the \ac{eft} depends on the presence of a charge density.
Clearly, it will be only valid within the cloud \( r < R_{\text{cl}}\). Near the edge of the cloud, there is however a region where the density is so low that it cannot be used anymore as the dominating scale.
In absence of the harmonic potential, the parameter \(\mu \) (or, equivalently its associated length scale \(R_\mu = \sqrt{2 \hbar/(m \mu)}\)) is the controlling parameter.
One possible physical interpretation for \(\mu\) is as a chemical potential or, equivalently, the result of the gauging with a flat connection.
This suggests defining an effective, position-dependent chemical potential in terms of the covariant derivative of \(\chi\) which, on-shell, coincides with \(U_0(r)\):
\begin{equation}
  \mu_{\text{eff}} (r) = D_t \chi = U_0(r).
\end{equation}
The \ac{eft} is valid as long as we probe length scales that are much bigger than the associated scale~\cite{Son:2005rv,Kravec:2018qnu}
\begin{equation}
  R_{\text{eff}}(r) = \sqrt{\frac{2 \hbar}{m \mu_{\text{eff}}(r)} } .
\end{equation}
Observing that the effective chemical potential at distances \(\delta \ll R_{\text{cl}}\) from the boundary is given by
\begin{equation}
  \mu_{eff}(R_{\text{cl}} - \delta) \sim \frac{m \omega^2}{\hbar} R_{\text{cl}} \delta,
\end{equation}
we can estimate the boundary of validity of the \ac{eft} as the distance from the origin such that \(R_{\text{eff}} \approx \delta\):
\begin{equation}
   R_{\text{eff}}(R_{\text{cl}} - \delta) \approx \delta \Rightarrow \delta \approx \sqrt{\frac{\hbar}{m} } \frac{1}{\mu^{1/6} \omega^{1/3}} \approx R_{\text{cl}}^{1/3} R_\mu^{2/3}.
\end{equation}
Equivalently, in terms of dimensionless quantities, the \ac{eft} is well-defined for \(0 < s < 1- \delta_s\), where
\begin{equation}\label{eq:delta_layer}
   1- \delta_s =  \frac{(R_{\text{cl}} - \delta)^2}{R_{\text{cl}}^2} \approx 1- 2 \frac{\delta}{R_{\text{cl}}} \Rightarrow \delta_s \approx \pqty{\frac{R_\mu}{R_{\text{cl}}} }^{2/3} = \pqty{\frac{\omega}{\mu} }^{2/3} = \epsilon^{2/3}.
\end{equation}
As expected, the validity of the \ac{eft} depends on the scale separation measured by the ratio
\begin{equation}
  \epsilon = \frac{R_\mu}{R_{\text{cl}}} \ll 1 .
\end{equation}
In the next section we will see that this is precisely the large-charge condition.
The different scales introduced above are represented in Figure~\ref{fig:scales}.

\medskip

In the following we will use the ratio \(\epsilon\) as the controlling parameter in the perturbative expansion.
In particular, the \ac{eom} for \(v\) now reads
\begin{equation}
	\left( \frac{v}{v_{LO}} \right)^\frac{4}{d} = (1 - s) + \epsilon^2 \frac{c_1}{c_0} \frac{s v'' + \frac{d}{2} v'}{v},
\end{equation}
where primes correspond to derivatives with respect to $s$, and
\begin{equation}
 v_{LO} = \sqrt{\hbar} \left[ \frac{4 c_0}{g R_\mu^2} \right]^\frac{d}{4}  
\end{equation}
is the \ac{vev} of the system without the harmonic potential (\emph{i.e.} $\omega \to 0$, $R_{cl} \to \infty$ and $s \to 0$).
The \ac{nlo} ground-state configuration that solves this equation is given by
\begin{equation} \label{eq:radial_vev}
	v = v_{LO} (1 - s)^\frac{d}{4} \left[ 1 - \epsilon^2 \frac{c_1}{c_0} \frac{d^2}{64} \left\{ \frac{4 - d}{(1 - s)^3} + \frac{3 d - 4}{(1 - s)^2} \right\} + \mathcal{O}(\epsilon^4) \right].
\end{equation}
As expected, the term $\propto c_1$ gives a semiclassical subleading correction. However, it diverges at the cloud edge $1-s \approx \delta_s$ because
\begin{equation}
  \frac{\epsilon^2}{(1-s)^3} \approx \frac{\epsilon^2}{\delta_s^3} \approx  1
\end{equation}
which shows explicitly that the perturbative expansion breaks, since the putative first-order term is comparable to the zeroth-order one. The breakdown of the bulk \ac{eft} means that one can only \emph{estimate} any divergent behavior occurring in this region by regularizing the integrals upon removing the $\delta_s$-layer. 

Physical quantities will have in general the same type of expansion in \(\epsilon\).
In the case of extensive quantities, obtained as integrals of densities, such as the charge or the energy, the dependence on \(\epsilon\) will come both from the perturbative expansion and from the \(\epsilon\)-dependence of the boundary of integration in the radial direction.

\subsubsection{The charge and the ground-state energy density}

Our goal is to express the physical quantities of our problem as perturbative expansions in terms of the inverse charge, which we consider fixed and small.
For ease of computation, it is however convenient to use a fixed chemical potential \(\mu\) and then express \(\mu\) as function of \(Q\) at the end.

\medskip
The ground-state charge density is related to the chemical potential $\mu$ through $\rho_0 = \frac{c_0}{\hbar} v^2$, \emph{i.e.}
\begin{equation}
	\rho_0 = \rho_{LO}  (1 - s)^\frac{d}{2} \left[ 1 - \epsilon^2 \frac{c_1}{c_0} \frac{d^2}{32} \left\{ \frac{4 - d}{(1 - s)^3} + \frac{3 d - 4}{(1 - s)^2} \right\} + \mathcal{O}(\epsilon^4) \right],
\end{equation}
where the leading-order term
\begin{equation}
  \rho_{LO} = c_0 \left[ \frac{2 c_0 m}{\hbar g} \mu \right]^{\frac{d}{2}}
\end{equation}
is the ground-state charge density without the harmonic trap. Integrating the charge density over the volume will tell us how the chemical potential is related to the charge $Q \gg 1$, although nonphysical divergences appear on the way and would need to be regulated.
More explicitly, we get
\begin{equation}
	Q = \frac{c_0 \Omega_d}{\hbar} \int_0^{R_{cl}} \dd{r} r^{d-1}  v(r)^2 = Q_{LO} \left[ 1 - I_{\text{div}} + \mathcal{O}(\epsilon^4) \right],
\end{equation}
where the leading term $Q_{LO}$ is defined through\footnote{The constant $\xi$ is introduced here in analogy with the notation of~\cite{Kravec:2018qnu}.}
\begin{equation}
	Q_{LO} = \frac{1}{(\xi  \epsilon)^d} = \frac{1}{\xi^d}  \left( \frac{\mu}{\omega} \right)^d,
\end{equation}
and $\xi = \sqrt{\frac{g}{4 \pi c_0}} \left[ \frac{2 \Gamma(d)}{c_0 \Gamma\left( \frac{d}{2} \right)} \right]^\frac{1}{d}$ is a constant of order one.
The divergent part $I_{\text{div}}$ that needs to be regularized is given by
\begin{equation} \label{eq:I_div}
	I_{\text{div}} = \epsilon^2 \frac{c_1}{c_0} \frac{d^2 \Gamma(d)}{16 \Gamma^2 \left( \frac{d}{2} \right)} \int_0^1 ds~s^{\frac{d}{2}-1} \left\{ \frac{4 - d}{(1 - s)^{3-\frac{d}{2}}} + \frac{3 d - 4}{(1 - s)^{2-\frac{d}{2}}} \right\}.
\end{equation}
One has to bear in mind that $Q$ is our physical control parameter and the above divergence has to be understood as a divergence in the expression of the chemical potential as a function of the charge.
Conveniently, the leading-order terms are not sensitive to \ac{ir} physics, which allows us to write a self-consistent \ac{eft}, dominated by a semiclassical configuration.
We can therefore conclude that
\begin{equation}
  \epsilon = \frac{\omega}{\mu} \sim Q^{-\frac{1}{d}}.
\end{equation}
This implies that the condition \( \epsilon = R_\mu/R_{\text{cl}} = \omega /\mu \ll 1\) \emph{is indeed equivalent to the large-charge condition $Q \gg 1$}. The parameter $\omega$ defining the strength of the harmonic potential can be regarded as fixing the units, so we should therefore think of $\mu \sim Q^\frac{1}{d}$ itself as a large parameter, just as in the relativistic case.

Finally, while the divergences can be cured by the addition of counterterms~\cite{Hellerman:2020eff}, here we will content ourselves with a qualitative description by merely removing the small $\delta_s$-layer (Eq.~\eqref{eq:delta_layer}) at the edge of the cloud from the domain of integration and estimate the \(Q\)-behavior of the contribution of the cutoff region.
The divergent integral Eq. (\ref{eq:I_div}) thus behaves like
\begin{equation}
	I_{\text{div}} \sim
	\begin{cases}
		\mathcal{O}(Q^{-\frac{2}{3}}) & d = 2, \\
		\mathcal{O}(Q^{-\frac{5}{9}}) & d = 3.
	\end{cases}
\end{equation}

\medskip

Similarly, we can compute the ground-state energy density. Because the \ac{vev} of the angular mode is space-independent, the expression Eq.~(\ref{energy_density}) of the energy density can be expressed as
\begin{equation}
	\mathcal{E}_0 = \frac{d}{d + 2} \frac{2 c_0 \hbar}{m R_\mu^2} v^2 \left[ \left( 1 + \epsilon^2 \frac{c_1}{c_0} \left( \frac{v'}{v} \right)^2 \right) \left( 1 + \frac{2}{d} \right)  s + \left( \frac{v}{v_{LO}} \right)^\frac{4}{d} \right],
\end{equation}
into which we can now plug the expression Eq.~(\ref{eq:radial_vev}) of $v$ to get
\begin{equation} \label{eq:E0_final}
	\mathcal{E}_0 = \mathcal{E}_{LO} (1 - s)^\frac{d}{2} \left[ 1 + \frac{2 s}{d} - \epsilon^2 \frac{c_1}{c_0} \frac{d (d + 2)}{32} \left\{ \frac{4 - d}{(1 - s)^3} + \frac{3 d - 4 - 2s}{(1 - s)^2} \right\} + \mathcal{O}(\epsilon^4) \right],
\end{equation}
where the leading-order term is given by
\begin{equation}
  \mathcal{E}_{LO} = \frac{d}{d + 2} \frac{g \hbar^2}{2 m} \left[ \frac{4 c_0}{g R_\mu^2} \right]^{\frac{d}{2}+1}.
\end{equation}
Here again, we stress that the leading-order term is not sensitive to the \ac{ir} behavior, which means the results can be trusted.
Integrating over the volume, we will obtain the conformal dimension of the lowest operator at large charge in the next section.

\subsubsection{Operator dimensions}

By virtue of the state-operator correspondence, the conformal dimension of the lowest operator at fixed charge is related to the total energy of the ground state with the harmonic trap by:
\begin{equation}
	\Delta = \frac{E_0}{\hbar \omega} = \frac{m}{2 \hbar^2} R_{\text{cl}} R_\mu  E_0, 
\end{equation}
where $E_0$ is the integral over the cloud of the ground-state energy density Eq.~(\ref{eq:E0_final}), and this yields the leading term of $\Delta$ to be of order $\epsilon^{-(d+1)} \sim Q^\frac{d+1}{d}$:
\begin{equation}
	\Delta = \frac{d}{d + 1} \xi Q^\frac{d+1}{d} + Q^\frac{d-1}{d}  \frac{c_1}{\xi c_0}  \frac{d^2 \Gamma(d)}{8 \Gamma^2\left( \frac{d}{2} \right)} \int_0^1 ds \frac{s^\frac{d}{2}}{(1 - s)^{2-\frac{d}{2}}}.
\end{equation}
The leading-order dependence of the conformal dimension on Q is thus exactly as in the relativistic case.
The second term is divergent in $d=2$ and turns into an $\mathcal{O}(Q^\frac{1}{2} \log Q)$ contribution when regulated, while the $d=3$ case is well-behaved. It should be noted, though, that our derivation does not allow for an explicit estimation of higher-order corrections because powers and/or logarithms of $Q$ can {\em a priori} arise when regularizing the diverging integrals, but it is known that the next contributions are of order $Q^\frac{5}{9}$ in $d=3$~\cite{Son:2005rv}. We therefore conclude that
\begin{equation}
  \label{eq:conformal-dimensions}
\Delta =
\begin{cases}
	\frac{2}{3} \xi Q^\frac{3}{2} + \mathcal{O}(Q^\frac{1}{2} \log Q)  & (d = 2) \\
	\frac{3}{4} \xi Q^\frac{4}{3} + \frac{27}{8 \xi} \frac{c_1}{c_0}  Q^\frac{2}{3} + \mathcal{O}(Q^\frac{5}{9}) & (d = 3).
\end{cases}
\end{equation}
The above results were first derived in~\cite{Kravec:2018qnu}, but using the \ac{nlsm} which was obtained via the coset construction. We on the other hand have worked exclusively with the \ac{lsm}.
The explicit appearance of the dilaton field in our formalism allows us to move away from the conformal point by introducing a small dilaton mass (see Section~\ref{sec:near-schroedinger}).
This would have been less transparent  in a \ac{nlsm} were the massive radial mode is integrated out.

\subsection{The Goldstone field}%
\label{sec:Golstones}

\subsubsection{The fluctuation spectrum}%
\label{sec:fluctuation-spectrum}

Now that we have computed the semiclassical contribution to the conformal dimensions, we can discuss the effect of the quantum fluctuations.
In order to derive the dispersion relations, let us consider small (normalized) fluctuations $\hat{\sigma}$ and $\hat{\chi}$ around the ground-state:
\begin{equation}
\begin{cases}
	\chi(t, \mathbf{x}) & = \mu  t + \frac{\hat{\chi}(t, \mathbf{x})}{\sqrt{c_0} v_{LO}}, \\
	f \sigma(t, \mathbf{x}) & = f \sigma_0(s) - \frac{\hat{\sigma}(t, \mathbf{x})}{\sqrt{c_0} v_{LO}},
\end{cases}
\end{equation}
where $f \sigma_0(s) = - \log(f v(s))$. In particular, the radial mode now reads $a(t, \mathbf{x}) = v(s) e^\frac{\hat{\sigma}(t, \mathbf{x})}{\sqrt{c_0} v_{LO}}$. By plugging these expressions into the Lagrangian (\ref{eq:LSMLagU}), we first get the constant measuring the leading effect, plus some terms linear in the fluctuations that are canceled on shell.
Next, we get the Lagrangian quadratic in the fluctuations, which is given (up to a boundary term) by
\begin{equation}
\begin{aligned}
	\mathcal{L}_{(2)} & = \left( \frac{v(s)}{v_{LO}} \right)^2 \left[ 2 \hat{\sigma} \dot{\hat{\chi}} - \frac{\hbar}{2 m} \left[ (\partial_i \hat{\chi})^2 + \frac{c_1}{c_0} (\partial_i \hat{\sigma})^2 \right] - \frac{1}{2} m_\sigma^2(s) \hat{\sigma}^2 \right].
\end{aligned}
\end{equation}
Due to the breaking of the $U(1)$, there is a space-dependent effective mass term $m_\sigma^2(s)$ for the dilaton, which is given by
\begin{equation}
	m^2_\sigma(s) = m^2_{LO} (1 - s) \left[1 - \epsilon^2 \frac{c_1}{c_0} \frac{d (d + 1)}{16} \left\{ \frac{4 - d}{(1 - s)^3} + \frac{3 d - 4}{(1 - s)^2} \right\} + \mathcal{O}(\epsilon^4) \right],
\end{equation}
where $m^2_{LO} = \frac{16 \hbar}{d m R_\mu^2}$ was obtained in~\cite{schroedinger} without the harmonic trap.
This shows explicitly that the scale at which the radial mode decouples is set by $R_\mu$. The \ac{eom} are given by
\begin{equation} \label{eq:LinEoM}
\begin{cases}
	\dot{\hat{\chi}} & = \frac{1}{2} m_\sigma^2(s) \hat{\sigma} - \frac{\hbar}{2 m} \frac{c_1}{c_0} \left[ \nabla^2 \hat{\sigma} + \frac{2 \del_i v}{v} \del_i \hat{\sigma} \right], \\
	\dot{\hat{\sigma}} & = \frac{\hbar}{2 m} \left[ \nabla^2 \hat{\chi} + \frac{2 \del_i v}{v} \del_i \hat{\chi} \right].
\end{cases}
\end{equation}
Since the system is linear and spherically invariant, we can look for a basis of  solutions separating the variables as 
\begin{equation} \label{eq:FluctDetailed}
\begin{aligned}
	\hat{\chi} & = e^{\left( \Lambda + \epsilon^2 \frac{c_1}{c_0} \lambda \right) i \omega t}  \left( F(s) + \epsilon^2 \frac{c_1}{c_0} f(s) \right)  Y_l,\\
	\hat{\sigma} & = e^{\left( \Lambda + \epsilon^2 \frac{c_1}{c_0} \lambda \right) i \omega t}  \epsilon \left( H(s) + \epsilon^2 \frac{c_1}{c_0} h(s) \right)  Y_l,
\end{aligned}
\end{equation}
where $Y_l$ is the spherical harmonic in $(d-1)$-dimensions, and $\Lambda$ and $\lambda$ have the interpretation of energies and are constants.
Note that the leading behavior is captured by the parameter $\Lambda$ and the functions $F(s)$ and $H(s)$, while subleading corrections are described by $\lambda$, $f(s)$ and $h(s)$. The linearized \ac{eom} Eq.~(\ref{eq:LinEoM}) can thus be separated into two parts. To order $\mathcal{O}(1)$, we get
\begin{equation} \label{eq:LinEoM_LO}
\begin{cases}
	i \Lambda F(s)  = \frac{4 (1 - s)}{d} H(s) \\
	i \Lambda H(s)  = s F''(s) + \frac{d (1 - 2 s)}{2 (1 - s)} F'(s) - \frac{l (l + d - 2)}{4 s} F(s),
\end{cases}
\end{equation}
while the $\mathcal{O}(\epsilon^2)$-pieces of the \ac{eom} is given by 
\begin{equation} \label{eq:LinEoM_NLO}
\begin{dcases}
  \begin{multlined}[b][.85\textwidth]
	i \lambda F(s) + i \Lambda f(s)  = \frac{4 (1 - s)}{d} h(s) - s H''(s) - \frac{d (1 - 2 s)}{2 (1 - s)} H'(s) \\
	 + \left[ \frac{l (l + d - 2)}{4 s} - \frac{d + 1}{4} \pqty{ \frac{4 - d}{(1 - s)^2} + \frac{3 d - 4}{1 - s} } \right] H(s),
  \end{multlined}
\\
\begin{multlined}[b][.85\textwidth]
	i \lambda H(s) + i \Lambda h(s)  = s f''(s) + \frac{d (1 - 2 s)}{2 (1 - s)} f'(s) - \frac{l (l + d - 2)}{4 s} f(s)\\
	 - \frac{d^2}{32} \pqty{ \frac{12 - 3 d}{(1 - s)^4} + \frac{6 d - 8}{(1 - s)^3} } s F'(s).
  \end{multlined}
\end{dcases}
\end{equation}
The two leading-order equations can be combined into
\begin{equation}\label{eq:hypergeom}
	0 = E_\Lambda(F) \equiv F''(s) + \frac{d (1 - 2 s)}{2 s (1 - s)} F'(s) + \left[ \frac{d \Lambda^2}{4 s (1 - s)} - \frac{l (l + d - 2)}{4 s^2} \right] F(s).
\end{equation}
This is a hypergeometric equation with a well-known solution.\footnote{See e.g.~\cite{Kravec:2018qnu}.}
Imposing regularity at the singular points $s = 0$ and $s = 1$ constrains the spectrum $\Lambda$ to take the form
\begin{equation}
  \label{eq:spectrum}
	\Lambda = \sqrt{\frac{4 n}{d} (n + l - 1) + 4 n + l},
\end{equation}
where $n = 0, 1, \ldots, \infty$, and $l\in \setZ$ for $d=2$ and $l \in \setN_0$ for $d>2$. Considering the equations for the \ac{nlo}, one finds
\begin{equation}
\begin{aligned}
	E_\Lambda(f) = & -\lambda \frac{d \Lambda}{2 s (1 - s)} F(s) + \frac{d^2}{32} \left( \frac{12 - 3 d}{(1 - s)^4} + \frac{6 d - 8}{(1 - s)^3} \right) F'(s) \\
	 & + \frac{i d \Lambda}{4 (1 - s)} \left[ H''(s) + \frac{d (1 - 2 s)}{2 s (1 - s)} H'(s) \right. \\
	 & + \left. \left\{ \frac{d + 1}{4 s} \left( \frac{4 - d}{(1 - s)^2} + \frac{3 d - 4}{1 - s} \right) - \frac{l (l + d - 2)}{4 s^2} \right\} H(s) \right].
\end{aligned}
\end{equation}
This equation admits regular solutions if and only if the right-hand side has poles of order at most one in $s=0,1$.
However, around \(s =1\) we find
\begin{equation}
  E_\Lambda(f) = \frac{d \left( F(1) \left(d^3+d^2-4 d-16\right) \Lambda ^2 - 6d (d-4) F'(1)  \right)}{64 (s-1)^4} + \order{\frac{1}{ (s-1)^3}} ,
\end{equation}
and since \(F\) and \(F'\) are regular and non-vanishing at the edge, we see that there is a pole of order four in $s=1$.
This is again a manifestation of the edge singularity which needs to be renormalized.
Without the explicit form of the counterterm we can only say that the spectrum will receive a correction at $\order{\epsilon^2}$.

\subsubsection{Casimir energy}%
\label{sec:Casimir}

The first quantum correction to the semiclassical result comes from the Casimir energy of the fluctuations over the fixed-charge ground state.
Using the Coleman--Weinberg formula and the spectrum in Eq.~\eqref{eq:spectrum} we find
\begin{equation}
  E_{\text{Cas}} = \frac{\omega}{2} \sum_{n,l} \sqrt{ \frac{4 n}{d} \pqty{n + l - 1} + 4n + l } .
\end{equation}
This series is clearly divergent and needs to be regularized.
One can use for example the zeta-function regularization and write (for $d = 2$):
\begin{equation}
  E^{(d=2)}_{\text{Cas}} = \omega \eval{E_2(s)}_{s=-\frac{1}{2}} = \frac{\omega}{2} \eval{ \sum_{\substack{n=0,l = -\infty \\ (n,l) \neq (0,0)}}^\infty  \pqty{ 2 \pqty{n + \frac{1}{2} } \pqty{n + \abs{l} + \frac{1}{2} } - \frac{1}{2} }^{-s}}_{s=-\frac{1}{2}}.
\end{equation}
The idea is to rewrite this expression in terms of a multivariate zeta function~\cite{Euler:1776}
\begin{equation}
  \zeta(s_1, s_2) = \sum_{n_1 > n_2 \ge 1 } \frac{1}{n_1^{s_1} n_2^{s_2}} 
\end{equation}
that can be analytically continued to a meromorphic function on \(\setC^2\)~\cite{Goncharov:2001iea}.

\begin{figure}
  \centering
  \tikzset{mystyle/.style={shape=circle,fill=black,scale=0.3}}
\tikzset{withtext/.style={fill=white}}

  \begin{tikzpicture}[scale=.5]
    \foreach \x in {0,...,8}
    \foreach \y in {1,...,4}
    {
      \node[mystyle] at (\x,\y){};
    }
    \foreach \x in {0,...,8}
    \foreach \y in {-4,...,-1}
    {
      \node[mystyle] at (\x,\y){};
    }
    \foreach \x in {1,...,8}
    {
      \node[mystyle] at (\x,0){};
    }
    \draw[-latex] (-.5,0) -- (8.5,0);
    \node at (9,0) {\(n\)};
    \draw[-latex] (0,-4.5) -- (0,4.5);
    \node at  (0,5) {\(l\)};
    \draw[smooth,red,thick] (.75,4.5) -- (.75,.75) -- (8.5, .75);
    \node[red] at (-.75,4.5) {\(I\)};
    \draw[smooth,red,thick] (.25,4.5) -- (.25,.75) -- (-.25, .75) -- (-.25,4.5);
    \node[red] at (8.5,4.5) {\(II\)};
    \draw[smooth,red,thick] (8.5,.25) -- (.75,.25) -- (.75, -.25) -- (8.5,-.25);
    \node[red] at (8.75,-.75) {\(III\)};
  \end{tikzpicture}
  \caption{Decomposition of the summing region. The summand is symmetric under the exchange \(l \to - l\).}
  \label{fig:summing-regions}
\end{figure}
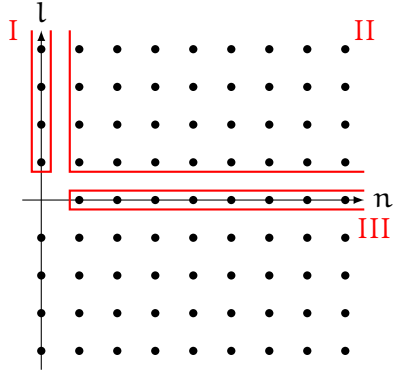

First we use the binomial expansion
\begin{equation}
  E_2(s) = \frac{1}{2^{s+1}} \sum_{n,l} \sum_{k=0}^{\infty} \binom{-s}{k} \pqty{-\frac{1}{4} }^k \pqty{n + \frac{1}{2} }^{-s-k} \pqty{n + l + \frac{1}{2} }^{-s -k} = \sum_{n,l,k} e_{n,l,k}(s)
\end{equation}
and separate the summing region into three parts (see figure~\ref{fig:summing-regions}):
\begin{align}
  I &= \set{(n,l) | n = 0, l \in \setN} \\
  II &= \set{ (n,l) | n \in \setN , l \in \setN } \\
  III &= \set{ (n,l) | n \in \setN, l = 0}.
\end{align}
The sum becomes
\begin{equation}
  E_2(s) = 2 E_2^I (s) + 2 E_2^{II} (s) + E_2^{III}(s),
\end{equation}
where
\begin{align}
  E_2^I(s) &= \sum_{l=1}^\infty \sum_{k = 0}^\infty e_{n,l,k}(s) = \frac{1}{2^{s+1}} \sum_{k=0}^\infty \binom{-s}{k}(-1)^k 2^{s - k} \zeta_H(s +k| \tfrac{3}{2}), \\
  E_2^{III}(s) &= \sum_{n=1}^\infty \sum_{k = 0}^\infty e_{n,l,k}(s) = \frac{1}{2^{s+1}} \sum_{k=0}^\infty \binom{-s}{k}\pqty{- \frac{1}{4} }^k  \zeta_H(2s +2k| \tfrac{3}{2})
\end{align}
and \(\zeta_H(s|x)\) is the Hurwitz zeta function
\begin{equation}
  \zeta_H(s| x) = \sum_{n = 0}^\infty \frac{1}{(s + x)^s} = \sum_{k=0}^\infty \binom{-s}{k} \pqty{x - 1}^{k} \zeta(s+k) .
\end{equation}
In the region \(II = \set{n > 1, l >1}\) we can write the sum in terms of the multivariate Hurwitz zeta function~\cite{murty2006multiple}
\begin{equation}
  \zeta_H(s_1, s_2 | x_1, x_2) = \sum_{n_1 > n_2 \ge 1 } \frac{1}{(n_1 + x_1)^{s_1} (n_2 + x_2)^{s_2}} 
\end{equation}
as
\begin{equation}
  E_2^{II}(s) = \frac{1}{2^{s+1}} \sum_{k=0}^\infty \binom{-s}{k} \pqty{-\frac{1}{4} }^k \zeta_H(s+k, s+k| \tfrac{1}{2},\tfrac{1}{2})
\end{equation}
or then apply the binomial expansion twice and write
\begin{equation}
  E_2^{II}(s) = \frac{1}{2^{s+1}} \sum_{k, j_1, j_2 = 0}^\infty \binom{-s}{k} \pqty{-\frac{1}{4} }^k \binom{-s-k}{j_1} \binom{-s-k}{j_2} \pqty{\frac{1}{2}}^{j_1 + j_2} \zeta(s +k +j_2, s+k +j_1) .
\end{equation}
The factor in front of the zeta function in the sum is symmetric under the exchange \(j_1 \leftrightarrow j_2\), so we can use the reflection identity~\cite{Euler:1776}
\begin{equation}
  \zeta(s_1, s_2) + \zeta(s_2, s_1) = \zeta(s_1) \zeta(s_2) - \zeta(s_1 + s_2)
\end{equation}
and rewrite \(E_2^{II}(s)\) as
\begin{multline}
  E_2^{II}(s) =  \frac{1}{2^{s+2}} \sum_{k = 0}^\infty \binom{-s}{k}  \pqty{-\frac{1}{4} }^k \Bigg[ \zeta_H(s+k|\tfrac{3}{2})^2 \\ - \sum_{j_1,j_2 = 0 }^\infty \binom{-s-k}{j_1} \binom{-s-k}{j_2} \frac{1}{2^{j_1 + j_2}} \zeta(2s +2 k +j_1 + j_2) \Bigg] . 
\end{multline}
The sums in \(E_2^{II}(s)\) and \(E_2^{III}(s)\) have poles in \(s = -1/2\), respectively with residue \(\sqrt{2}/16\) and \(-\sqrt{2}/8\), because they both include \(\zeta(1)\).
The fact that the two poles cancel in \(E_2(s)\) is a nice confirmation of our chosen regularization.
The final series can be evaluated numerically, since it converges very rapidly, to find
\begin{equation}
  E^{(d=2)}_{\text{Cas}} = \omega E_2(-\tfrac{1}{2}) = -0.294159\dots \times \omega .
\end{equation}
Note that this contribution, which is \(Q\)-independent, is parametrically smaller than the estimated behavior of the boundary term in  Eq.~\eqref{eq:conformal-dimensions}.
The same approach can be used in higher dimensions, however, generically the zeta function has a pole at $s=-1/2$ which would need to be regularized with appropriate counterterms.

\section{Near-Schrödinger dynamics at large charge}%
\label{sec:near-schroedinger}

After carefully developing the large-charge results for the Schrödinger particle in the harmonic trap, we now extend our treatment to the near-Schrödinger regime by giving a small mass to the dilaton which explicitly breaks Schrödinger invariance.

\subsection{Explicit breaking of Schrödinger invariance}

We first present the general construction of a potential for the dilaton that explicitly breaks scale invariance as first proposed by Coleman~\cite{Coleman:1988aos}.
In a relativistic ($z = 1$) or non-relativistic ($z \neq 1$) theory in $(d+1)$-dimensions, the dilaton transforms non-linearly under scale transformations as
\begin{equation}
	\sigma(t, x_i)  \rightarrow \sigma(e^{2\tau}t, e^{\tau} x_i) - \frac{d + z - 2}{2 f} \tau,
\end{equation}
where $f$ is a constant of dimension $[f^{-1}] = [\sigma] = [a]$, and $a \sim \frac{1}{f} e^{-f \sigma}$ corresponds to the \emph{radial mode} and is canonically normalized. As discussed in Section~\ref{sec:schroedinger}, any local operator can be made scale-invariant by dressing it with an appropriate power of the dilaton.
In particular, a constant operator is dressed as $c \to c e^{-2 \eta f \sigma}$, where
\begin{equation}
	\eta = \frac{d + z}{d + z - 2}.
\end{equation}
This gives a scale-invariant potential for the dilaton (or, equivalently, for the radial mode).
Coleman's potential consists of such a term whose constant $c$ is small (in a sense that is explained below) and an additional scale-symmetry breaking piece which gives rise to a mass term for $\sigma$, namely
\begin{equation}
\begin{aligned}
	U_C & = \frac{m_\sigma^2}{4 \eta^2 f^2} \left[ e^{-2 \eta f \sigma} + 2 \eta f \sigma - 1 \right] \\
	& = \frac{m_\sigma^2}{4 \eta^2 f^2} \left[ (f a)^{2 \eta} - 2 \eta \log(f a) - 1 \right].
\end{aligned}
\end{equation}
$m_\sigma$ is a small parameter that we will refer to as the \emph{dilaton mass}, although $[m_\sigma^2] = T^{-1}$ in the non-relativistic case.
In fact, to quadratic order, 
\begin{equation}
	U_C \approx \frac{1}{2} m_\sigma^2 \sigma^2.
\end{equation}
The linear piece in $\sigma$ is needed to eliminate the tadpole hidden in the exponential, and ends up giving the most important contribution. This will play an important role in Section~\ref{sec:Golstones_dilaton}.

\medskip
In order to understand in what sense $m_\sigma$ needs to be small, consider an effective theory that originally contains a scale-invariant potential of the form $\lambda u a^{2 \eta}$, where $\lambda \sim \mathcal{O}(1)$ is a Wilsonian coefficient and $u$ keeps track of factors of $\hbar$ and others, if needed. The addition of Coleman's potential $U_C$ should be a small deviation from the original theory, which means we {should fine tune} 
\begin{equation}
	\frac{f^{2(\eta - 1)}}{u} m_\sigma^2 \ll 1.
\end{equation}

\medskip
Finally, let us comment on the trace of the stress tensor. Although the breaking of scale invariance is very explicit due to the last two terms in Coleman's potential, we can actually quantify {\em how much} we break it. Indeed, classically, the trace of the stress tensor does no longer vanish but equals\footnote{For more details on the tracelessness of the stress tensor in non-relativistic theories, we refer the reader to \cite{Hagen:1972pd}.}
\begin{equation}
		T = \frac{d + z - 2}{2} \frac{m^2_\sigma}{f} \sigma.
\end{equation}

\subsection{Semiclassical analysis}

We now specialize to $z = 2$ and $\eta = \frac{d + 2}{d}$, and we couple the Lagrangian (\ref{eq:LSMLagU}) to Coleman's potential,
\begin{equation} \label{eq:LSMLagDilaton}
	\tilde{\mathcal{L}} = \mathcal{L} - c_0 U_C.
\end{equation}
Let us study the scales involved in our problem. First, we are still looking for the \ac{nlo} contributions of the large-charge expansions of the observables. Typically, we want to write the ground-state solution up to $\order{\epsilon^4}$ and compute the corresponding energy. Next, we associate length scales to the parameters $f$ and $m_\sigma$, respectively:
\begin{equation}
	R_f = \left( \hbar f^2 \right)^\frac{1}{d}
	\hspace{1cm} \mathrm{and} \hspace{1cm}
	R_\sigma = \sqrt{\frac{2 \hbar}{m m_\sigma^2}}.
\end{equation}
In terms of those, the condition for $m_\sigma$ being small simply reads $R_\sigma \gg R_f$, and we thus define a new small expansion parameter
\begin{equation}
	\tilde{\epsilon} = \frac{R_f}{R_\sigma} \propto m_\sigma
\end{equation}
In what follows, we shall be merely interested in the signature {\em linear in the small dilaton mass} in the observables. This, in turn, means that we will only keep track of terms of order $\tilde{\epsilon}^2$. Note that such corrections may affect the leading and subleading terms of the large-charge expansion.

\medskip
Going back to Eq.~\eqref{eq:LSMLagDilaton}, we can now investigate how the new potential modifies our previous results. We first note that the charge density is still given by $\rho = \frac{c_0}{\hbar} a^2$ and the continuity equation is unaffected, but the radial mode now has to satisfy the new \ac{eom}
\begin{equation}\label{eq:eom_bis}
	\left[ 1 + \tilde{\epsilon}^2 \frac{d c_0}{(d + 2) g} \right] a^{\frac{4}{d}+2} = \frac{2 m c_0 \hbar^\frac{2}{d}}{g \hbar} \left[ a^2 U + \frac{c_1}{c_0} \frac{\hbar}{2 m} a \nabla^2 a \right] + \frac{d \hbar^{\frac{2}{d}+1}}{(d + 2) g} \frac{c_0 \tilde{\epsilon}^2}{R_f^{d+2}}.
\end{equation}
Moreover, the energy density is simply given by
\begin{equation}
	\tilde{\mathcal{E}} = \mathcal{E} + c_0 U_C.
\end{equation}

\subsubsection{The ground-state solution}

The identification of the ground-state solution $(\tilde{v}, \chi_0)$ for the radial and the angular modes respectively follows along the same line as before. Again, the limited number of \ac{dof} prevents spherical invariance from being broken on top of dilatations and $U(1)$, and a minimal-energy solution to the \ac{eom} is still given by $\chi_0 = \mu t$ and a time-independent \ac{vev} $\tilde{v}$ for the radial mode. The latter, however, receives corrections in order to solve the modified \ac{eom} Eq. (\ref{eq:eom_bis}) that can be written here as
\begin{equation}
	\left( \frac{\tilde{v}}{\tilde{v}_{LO}} \right)^\frac{4}{d} = (1 - s) + \epsilon^2 \frac{c_1}{c_0} \frac{s \tilde{v}'' + \frac{d}{2} \tilde{v}'}{\tilde{v}} + \left( \frac{\epsilon}{\alpha} \right)^{\frac{d}{2}+1}  \frac{c_0 \tilde{\epsilon}^2}{g} \frac{d}{d+2} \left( \frac{\tilde{v}_{LO}}{\tilde{v}} \right)^2 + \mathcal{O}(\tilde{\epsilon}^4),
\end{equation}
where we have defined the constant $\alpha = \frac{2 m \omega c_0 R_f^2}{g \hbar}$. Moreover, the leading-order coefficient is redefined as 
\begin{equation}
	\tilde{v}_{LO} = v_{LO} \left( 1 - \frac{c_0 \tilde{\epsilon}^2}{g} \frac{d^2}{4 (d + 2)} \right).
\end{equation}
The truncated solution reads
\begin{equation} \label{eq:radial_vev_dilaton}
\begin{aligned}
	\tilde{v} = \tilde{v}_{LO} (1 - s)^\frac{d}{4} & \left[ 1 - \epsilon^2 \frac{c_1}{c_0} \frac{d^2}{64} \left\{ \frac{4 - d}{(1 - s)^3} + \frac{3 d - 4}{(1 - s)^2} \right\} \right. \\
	& \left. \hspace{5mm} + \left( \frac{\epsilon}{\alpha (1 - s)} \right)^{\frac{d}{2}+1}  \frac{c_0 \tilde{\epsilon}^2}{g} \frac{d^2}{4 (d + 2)} + \mathcal{O}(\epsilon^4) \right],
\end{aligned}
\end{equation}
where $\mathcal{O}(\epsilon^4)$ is a shorthand notation for higher-order contributions of both $\epsilon$ and $\tilde{\epsilon}$. The last term is well-behaved under the $\delta_s$-regularization that we introduced earlier, so that no modification is needed when investigating the boundary behavior of the theory.
It is important to note that the qualitative description of the breaking of $U(1)$ is unaffected as the \ac{vev} scales like $\mu^{d/4}$.

\subsubsection{The charge and the ground-state energy density}

Now that we have the correction to the ground state due to the presence of Coleman's potential, let us investigate how the chemical potential $\mu$ is related to the total charge $Q = \frac{c_0}{\hbar} \int \dd^d x\, \tilde{v}^2$. We find that
\begin{equation}
	Q = \tilde{Q}_{LO} \left[ 1 - I_{div} + \tilde{I}_{div} \right],
\end{equation}
where the leading order is given by $Q_{LO} = (\tilde{\xi}  \epsilon)^{-d}$, similarly to before, except for the modified parameter 
\begin{equation}
	\tilde{\xi} = \xi \left( 1 + \frac{c_0 \tilde{\epsilon}^2}{g} \frac{d}{2 (d + 2)} \right).
\end{equation}
Moreover, there is yet another divergent part that, after reguralization of the $\delta_s$-layer, is given by
\begin{equation}
	\tilde{I}_{\text{div}} = \left( \frac{\epsilon}{\alpha} \right)^{\frac{d}{2}+1}  \frac{c_0 \tilde{\epsilon}^2}{g} \frac{d^2 \Gamma(d)}{(d + 2) \Gamma^2 \left( \frac{d}{2} \right)} \int_0^{1-\delta_s} ds~\frac{s^{\frac{d}{2}-1}}{(1 - s)} \sim \mathcal{O}(Q^{-\frac{d+2}{2d}} \log(Q)).
\end{equation}

\medskip

We can now turn to the computation of the ground-state energy density in order to, then, express the conformal dimension in terms of the charge. We have
\begin{equation} \label{eq:GS_energy_density_dilaton}
\begin{aligned}
	\tilde{\mathcal{E}}_0 = \tilde{\mathcal{E}}_{LO} & \left[ \left( 1 + \frac{2 s}{d} \right) (1 - s)^\frac{d}{2} - \epsilon^2 \frac{c_1}{c_0} \frac{d (d + 2)}{32} \left\{ \frac{4 - d}{(1- s)^{3-\frac{d}{2}}} + \frac{3 d - 4 - 2 s}{(1 - s)^{2-\frac{d}{2}}} \right\} \right. \\ 
	& \left. + \left( \frac{\epsilon}{\alpha} \right)^{\frac{d}{2}+1}  \frac{c_0 \tilde{\epsilon}^2}{g} \frac{d}{2} \left\{ \frac{1}{1 - s} + \log \epsilon - \log[\alpha (1 - s)] - \frac{2}{d + 2} \right\}+\order{\epsilon^4} \right].
\end{aligned}
\end{equation}
Note that the leading coefficient is now given by $\tilde{\mathcal{E}}_{LO} = \mathcal{E}_{LO} \left( 1 - \frac{c_0 \tilde{\epsilon}^2}{g} \frac{d^2}{2 (d + 2)} \right)$.
The presence of a $\log \epsilon$ term that will eventually constitute the most important contribution of Coleman's potential to the conformal dimension.

\subsubsection{Operator dimensions}%
\label{sec:observables_dilaton}

As usual, the physical observable of interest is the scaling dimension $\tilde{\Delta}$ of the lowest operator in the near-Schrödinger framework.
Note that, strictly speaking, the scaling dimension is not defined for broken Schrödinger symmetry, but since we assume the breaking to small, we still assume that the two-point function has a power-law behavior and that we can estimate the putative conformal dimensions using the state-operator correspondence:
\begin{equation}
	\tilde{\Delta} = \frac{m}{2 \hbar^2} R_{\text{cl}} R_\mu  \tilde{E}_0,
\end{equation}
where the lowest energy level $\tilde{E}_0$ is found  integrating the ground-state energy density over the cloud. As anticipated, Eq.~(\ref{eq:GS_energy_density_dilaton}) indicates the presence of a logarithmic contribution of the charge to the conformal dimension. One major difference to the relativistic case, though, is the explicit dependence of the domain of integration on the charge: the volume of the cloud scales like $\sqrt{Q}$, indicating that the bigger the charge (particle number), the larger the cloud. The remaining terms in the energy are just constants multiplying the volume $\sqrt{Q}$, and we cannot see these contributions neither in $d=2$ nor in $d=3$. In the end, we have
\begin{multline}
	\tilde{\Delta} = \frac{d}{d + 1} \tilde{\xi} Q^\frac{d+1}{d} + Q^\frac{d-1}{d}  \frac{c_1}{\tilde{\xi} c_0}  \frac{d^2 \Gamma(d)}{8 \Gamma^2\left( \frac{d}{2} \right)} \int_0^1 ds \frac{s^\frac{d}{2}}{(1 - s)^{2-\frac{d}{2}}}\\
	 - \sqrt{Q} \log Q  \frac{\tilde{\epsilon}^2}{\alpha^{\frac{d}{2}+1}} \frac{c_0}{g \tilde{\xi}^\frac{d}{2}} \frac{2 \Gamma(d)}{(d + 2) \Gamma^2 \left( \frac{d}{2} \right)},
\end{multline}
which translates into%
\begin{equation}\label{eq:conformal-dimensions-broken}
\tilde{\Delta} = 
\begin{cases}
	\frac{2}{3} \xi Q^\frac{3}{2} + \mathcal{O}(Q^\frac{1}{2} \log Q)  & (d = 2), \\
	\frac{3}{4} \tilde{\xi} Q^\frac{4}{3} + \frac{27}{8 \tilde{\xi}} \frac{c_1}{c_0}  Q^\frac{2}{3} - \frac{\kappa m_\sigma^2}{\tilde{\xi}^{3/2}} \sqrt{Q} \log(Q) + \mathcal{O}(Q^{\frac{5}{9}}) & (d = 3),
\end{cases}
\end{equation}
where $\kappa = \frac{\tilde{\epsilon}^2}{\alpha^\frac{5}{2}} \frac{c_0}{g m_\sigma^2} \frac{16}{5 \pi}$.
This is a purely semi-classical result. The divergence is due only to the \ac{nlo} term while the dilaton mass contribution does not need to be regularized. We will see that once we consider also the quantum effects, this is no longer the case. 

In \(d=2\), the leading effect of the breaking of Schrödinger invariance is precisely of the same order as the expected contribution of the edge singularity and is in this sense undetectable within the limits of this analysis.
In \(d = 3\) the situation is more involved.
Assuming that all the coefficients are of order one or smaller, there is a large interval of values of the charge where the \(Q \log(Q)\) associated to the dilaton mass dominates over the edge effects that scale like \(Q^{5/9}\) (numerically, \(Q \log(Q) > Q^{5/9}\) over \(34\) orders of magnitude).
This suggests that it might still be possible to measure the effect of the breaking of the Schrödinger symmetry from an independent computation of \(\Delta(Q)\) \emph{e.g.} on the lattice.

\subsection{The Goldstone field}\label{sec:Golstones_dilaton}

In order to determine how the Goldstone dynamics is influenced by the presence of the small dilaton mass, let us repeat the steps of Section~\ref{sec:Golstones}.
Normalizing the fluctuations as
\begin{equation}
\begin{cases}
	\chi(t, \mathbf{x}) & = \mu  t + \frac{\hat{\chi}(t, \mathbf{x})}{\sqrt{c_0} \tilde{v}_{LO}}, \\
	f \sigma(t, \mathbf{x}) & = f \tilde{\sigma}_0(s) - \frac{\hat{\sigma}(t, \mathbf{x})}{\sqrt{c_0} \tilde{v}_{LO}},
\end{cases}
\end{equation}
where $f \tilde{\sigma}_0(s) = -\log(f \tilde{v}(s))$ is the deformed \ac{vev} of the dilaton, we get the Lagrangian quadratic in the fluctuations:
\begin{equation}
\begin{aligned}
	\mathcal{L}_{(2)} & = \left( \frac{\tilde{v}(s)}{\tilde{v}_{LO}} \right)^2 \left[ 2 \hat{\sigma} \dot{\hat{\chi}} - \frac{\hbar}{2 m} \left[ (\partial_i \hat{\chi})^2 + \frac{c_1}{c_0} (\partial_i \hat{\sigma})^2 \right] - \frac{1}{2} \tilde{m}_\sigma^2(s) \hat{\sigma}^2 \right].
\end{aligned}
\end{equation}
This is almost completely the same as before, except for the important modification of the mass term:
\begin{equation}
\begin{aligned}
	m^2_\sigma(s) = m^2_{LO} (1 - s) & \left[1 - \epsilon^2 \frac{c_1}{c_0} \frac{d (d + 1)}{16} \left\{ \frac{4 - d}{(1 - s)^3} + \frac{3 d - 4}{(1 - s)^2} \right\} \right. \\
	& \left. \hspace{5mm} + \left( \frac{\epsilon}{\alpha (1 - s)} \right)^{\frac{d}{2}+1}  \frac{c_0 \tilde{\epsilon}^2}{g} \frac{d}{2} + \mathcal{O}(\epsilon^4) \right].
\end{aligned}
\end{equation}
The last term comes directly from Coleman's potential. The \ac{eom} are exactly the same as before with $v(s)$ replaced by the deformed \ac{vev} $\tilde{v}(s)$, and the correction due to the dilaton mass can be captured  expanding the fluctuations as
\begin{equation} \label{eq:FluctDetailed2}
\begin{aligned}
	\hat{\chi} & = e^{\left( \Lambda + \left( \frac{\epsilon}{\alpha} \right)^{\frac{d}{2}+1} \frac{c_0 \tilde{\epsilon}^2}{g} \tilde{\lambda} \right) i \omega t}  \left( F(s) + \left( \frac{\epsilon}{\alpha} \right)^{\frac{d}{2}+1} \frac{c_0 \tilde{\epsilon}^2}{g} \tilde{f}(s) \right)  Y_l, \\
	\hat{\sigma} & = e^{\left( \Lambda + \left( \frac{\epsilon}{\alpha} \right)^{\frac{d}{2}+1} \frac{c_0 \tilde{\epsilon}^2}{g} \tilde{\lambda} \right) i \omega t}  \epsilon \left( H(s) + \left( \frac{\epsilon}{\alpha} \right)^{\frac{d}{2}+1} \frac{c_0 \tilde{\epsilon}^2}{g} \tilde{h}(s) \right)  Y_l,
\end{aligned}
\end{equation}
where $\tilde{\lambda}$ encodes the correction to the leading-order spectrum found in Section \ref{sec:Golstones}. This spectrum is constrained by the $\mathcal{O}\left(\epsilon^{\frac{d}{2}+1}\right)$-pieces of the \ac{eom}, which read
\begin{equation} \label{eq:LinEoM_NLODilaton}
\begin{cases}
	i \tilde{\lambda} F(s) + i \Lambda \tilde{f}(s) & = \frac{4 (1 - s)}{d} \tilde{h}(s) + \frac{2}{(1 - s)^\frac{d}{2}} H(s) \\
	i \tilde{\lambda} H(s) + i \Lambda \tilde{h}(s) & = s \tilde{f}''(s) + \frac{d (1 - 2 s)}{2 (1 - s)} \tilde{f}'(s) - \frac{l (l + d - 2)}{4 s} \tilde{f}(s) + \frac{d^2}{4} \frac{s}{(1 - s)^{\frac{d}{2}+2}} F'(s).
\end{cases}
\end{equation}
They can be combined into
\begin{equation}
	E_\Lambda(\tilde{f}) = \frac{d \Lambda}{2 s (1 - s)} \left[ \frac{d \Lambda}{4 (1 - s)^{\frac{d}{2}+1}} - \tilde{\lambda} \right] F(s) - \frac{d^2}{4} \frac{1}{(1 - s)^{\frac{d}{2}+2}} F'(s).
\end{equation}
Substituting the explicit solution (which is regular and non-vanishing at \(s =1\)), we find that the right-hand side has a pole of order $2+d/2$ at $s=1$ and there are no regular solutions.
This means that we need to regularize also the contribution from the dilaton mass to the spectrum to account for the edge singularity.

\section{Conclusions and Outlook}%
\label{sec:conclusions}

Non-relativistic systems with Schrödinger symmetry have many parallels to relativistic \acp{cft}. In both cases, the symmetry gives rise to constraints on correlation functions and there is a precise notion of a state-operator correspondence.

\medskip
In this paper, we have shown first, that it is possible to construct a fully scale-invariant theory from a Galilean-invariant one by introducing a dilaton field and dressing all operators with appropriate powers of it. This procedure is an exact parallel of Coleman's dilaton dressing for the relativistic case. 

\medskip
The large-charge approach is useful for treating strongly coupled theories which have otherwise no small parameters. While in principle, the approach is not tied to any particular space-time symmetry, working at the conformal point has many practical advantages. Schrödinger symmetry has proven to be a suitable setting for working at large charge as well, even though the need to work with a harmonic potential in order to use the state-operator correspondence introduces considerable complications due to the breakdown of the bulk \ac{eft} near the edge of the particle cloud, which has only recently been addressed in~\cite{Hellerman:2020eff}.

We have reproduced results for the Schrödinger system at large charge in the harmonic potential, but in the \ac{lsm} in order to then move away slightly from the case of full Schrödinger symmetry, further expanding the scope of applicability of the large-charge approach to more general systems.

Using the description of the radial mode in terms of the dilaton, we introduced a small explicit breaking of Schrödinger symmetry by including a potential for the dilaton mimicking Coleman's mass term. This allowed us to compute the corrections to the system at large charge with near-Schrödinger dynamics. We moreover gave the result for the Casimir energy of the fluctuations over the ground state which had not been computed for the Schrödinger system to date.

This explicit breaking of Schrödinger symmetry results in a correction to the scaling dimension with a signature term that scales as $\sqrt{Q}\log{Q}$. While in $d=2$, it is of the same oder as the contribution of the edge effects, in $d=3$, it dominates the edge effects over a range of 34 orders of magnitude and should be discernible in lattice simulations.

\medskip
There are a number of open questions for future study.
Since the unitary Fermi gas can be realized in the lab by tuning Feshbach resonances, it would be interesting to understand if the breaking via the dilaton mass discussed here is related to a real-life tuning process that can happen in the lab. In this spirit, near-unitarity of the Fermi gas was for example discussed in~\cite{Escobedo:2009bh}.

A constant obstacle to our analysis here were the unknown contributions from the cloud edge, forcing us to employ estimates which often appear to mask the contributions we are interested in. The near-Schrödinger case discussed here should thus be revisited employing the edge \ac{eft} of~\cite{Hellerman:2020eff} which would allow us to sharpen the signal of the Schrödinger-symmetry breaking term.

\subsection*{Acknowledgements}

 We would like to thank Simeon Hellerman for discussions of the edge \ac{eft} and sharing some unpublished results.
 The work of S.R. and V.P. is supported by the Swiss National Science Foundation under grant number 200021 192137.
 D.O. acknowledges partial support by the \textsc{nccr 51nf40--141869} ``The Mathematics of Physics'' (Swiss\textsc{map}).

\setstretch{1}

\printbibliography{}

\end{document}